\documentclass[aps,prl,twocolumn,superscriptaddress,floatfix]{revtex4-2}

\usepackage{amsmath,amssymb,amsfonts}
\usepackage{graphicx}
\usepackage{hyperref}
\usepackage{bm}
\usepackage{placeins}

\newcommand{\Chtwo}{\mathrm{Ch}_{2}}
\newcommand{\Chone}{\mathrm{Ch}_{1}}
\newcommand{\Wc}{W_{c}}
\newcommand{\Wg}{W_{g}}
\newcommand{\Choz}{\mathrm{Ch}_{0,3}}

\begin{document}

\title{Mobility-Gap Robustness and St\v{r}eda-Hall Separation in Disordered Axion
Pumping}

\author{Bryan Leung}
\email{bleung@spotify.com}
\affiliation{Spotify, New York, NY 10007, USA}

\author{Emil Prodan}
\email{prodan@yu.edu}
\affiliation{Department of Physics, Yeshiva University, New York, NY 10016, USA}

\date{\today}

\begin{abstract}
We show that the noncommutative second Chern number of a $(3{+}1)$D
disordered axion pump remains quantized beyond global spectral gap closure
at $\Wg\approx14.5$.  Large-scale computation of the second
Chern number and finite-size scaling demonstrate convergence of
$\Chtwo$ toward~$-1$
well into the mobility-gap regime ($W \lesssim 16.5$), while the
St\v{r}eda-Hall response $\partial_\phi \Choz$ departs rapidly after global
spectral gap closure, serving as an internal control that distinguishes bulk
mobility-gap protection from finite-size artifacts.  Level statistics
identify a 3D unitary Anderson transition at each pump
slice and locate the mobility-edge bottleneck at $\tau=0$, with
critical disorder $\Wc \approx 18$.
\end{abstract}

\maketitle

A three-dimensional (3D) axion insulator is characterized by a quantized
magnetoelectric coupling $\theta=\pi$ modulo $2\pi$, giving rise to
half-quantized Hall responses on symmetry-breaking gapped
surfaces~\cite{Qi2008,Essin2009,Mong2010,Leung2020}.
This quantization is protected by symmetries that send
$\theta\to-\theta$, such as time-reversal or inversion
symmetries, and remains robust against
moderate disorder~\cite{Leung2013,Prodan2013}.
Experiments have observed signatures of axion
electrodynamics~\cite{Wu2016,Mogi2017,Xiao2018,Liu2020,LiuGap2022,Zhuo2023},
and recent measurements report dynamical axion and half-quantized layer Hall
responses~\cite{Qiu2025,Hu2026}.
In contrast to a static axion insulator, an axion pump is an adiabatic cycle
$H(\tau)$ that generically breaks these symmetries, allowing
$\theta(\tau)$ to wind by $\Delta\theta=2\pi\Chtwo$, where $\Chtwo$ is
the second Chern number of the $(3{+}1)$D family.
Despite extensive study of static disordered axion
insulators~\cite{NomuraNagaosa2011,LeungProdan2012,Li2021,Song2021,Chen2025STI,Grindall2025}, the fate of dynamically pumped axion insulators
under disorder remains unexplored.

A central long-standing question is whether mobility-gap 
protection, which underlies static noncommutative Chern numbers
and integer quantum Hall plateaus, can also protect a dynamically
pumped Chern number after global spectral gap
closure~\cite{Bellissard1994,Prodan2013}. 
In one dimension ($1$D), quantized charge pumps under disorder
have been studied extensively~\cite{Thouless1983,Wauters2019,Cerjan2020,Hayward2021,Huang2025}.
The quantization of the pumped charge breaks down when
the spectral gap closes, even though the instantaneous
states remain localized~\cite{Hayward2021,SuppMat}.  As is well known, 
3D systems behave differently under disorder~\cite{Abrahams1979}. 
The spectrum does not localize entirely, as in $1$D, and instead mobility edges separate
localized and extended states.  Thus, even after global spectral gap
closure, states at $E_F$ can remain localized over a finite disorder
interval, leaving a mobility-gap regime in which noncommutative Chern
numbers may remain quantized~\cite{Prodan2013}.
Whether a disordered axion pump realizes the required spatiotemporal
regularity, and whether this is visible numerically, remain open.

The spectral-gap bulk-boundary
correspondence relates the second Chern number to surface spectral
flow~\cite{Leung2020},
and the associated generalized St\v{r}eda relation is the four-dimensional analogue of
the ordinary St\v{r}eda formula~\cite{ProdanSchulzBaldes2016,Streda1982}.  Instead of the change of charge density per unit
flux, it measures the change of the pumped weak Chern invariant $\Choz$ per unit
transverse flux, converting $\Chtwo$ into a topological magnetoelectric
response.  Because the identity assumes a spectral gap,
$\partial_\phi\Choz$ remains a useful control after global spectral gap
closure.  Thus, a separation between $\Chtwo$ and
$\partial_\phi\Choz$ after spectral gap closure distinguishes the
breakdown of the gap-dependent St\v{r}eda identity from mobility-gap
protection of $\Chtwo$.

We address these questions with a large-scale real-space computation
of the noncommutative second Chern number under strong disorder.
Independent scaling in system size and the kernel polynomial method (KPM)
order shows that $\Chtwo$ remains quantized beyond global spectral gap
closure, while the St\v{r}eda-Hall response $\partial_\phi\Choz$ departs rapidly
after global spectral gap closure.  Level statistics identify a
3D unitary Anderson transition at each pump slice, with the critical
disorder strength $\Wc=\min_\tau W_c(\tau)\approx18$ set by the $\tau=0$
mobility-edge bottleneck.  These results provide direct numerical evidence for
mobility-gap robustness of axion pumping. This is in stark contrast with the
$1$D disordered pump of Ref.~\cite{Hayward2021}, where our calculation of
the noncommutative first Chern number reproduces the
breakdown of pumped charge quantization after spectral gap
closure~\cite{SuppMat}.

We consider a clean $(3{+}1)$D axion pump model consisting of a stack of
bilayer Haldane sheets along~$z$ with $\tau$-dependent interlayer
coupling~\cite{Olsen2017},
\begin{align}
H_0(\tau) &= \sum_{\langle i,j \rangle} t_1 c_i^\dagger c_j
+ \sum_{\langle\langle i,j \rangle\rangle} i s_i \nu_{ij} t_2
  c_i^\dagger c_j \notag\\
&\quad + \frac{m(\tau)}{2} \sum_i s_i \sigma_i c_i^\dagger c_i
+ \sum_{\langle i,j \rangle_z} t^\perp_{ij}(\tau) \sigma_i
  c_i^\dagger c_j + \text{h.c.},
\label{eq:ham}
\end{align}
where $\langle i,j \rangle$ and $\langle\langle i,j \rangle\rangle$
denote nearest- and next-nearest-neighbor pairs on each layer, with
hopping sums taken over one orientation of each bond.  The Haldane
chirality factor is $\nu_{ij} = \pm 1$ (Haldane phase fixed at
$\pi/2$), and $t_1 = -4t_2$ with $t_2 = 1$ as the energy
unit.  The sublattice and layer labels are $\sigma_i = \pm 1$ and
$s_i = \pm 1$, and
$m(\tau) = (3\sqrt{3} + 2\cos\tau)\,t_2$.
The interlayer hopping is
$t^\perp_{ij}(\tau) = t_3^{\text{intra}}(\tau)\,\delta_{z_i,z_j}
+ t_3^{\text{inter}}(\tau)\,(1 - \delta_{z_i,z_j})$,
with $t_3^{\text{intra/inter}}(\tau) = (1 \pm \gamma \sin\tau)\,t_2$
and $\gamma = 0.4$.  Anderson disorder is added through the on-site
term $V_{\text{dis}} = \sum_i V_i c_i^\dagger c_i$, with $V_i$
drawn uniformly from $[-W/2, W/2]$, giving the full
Hamiltonian $H(\tau)=H_0(\tau) + V_{\text{dis}}$.

\begin{figure}[!t]
\centering
\includegraphics[width=\columnwidth]{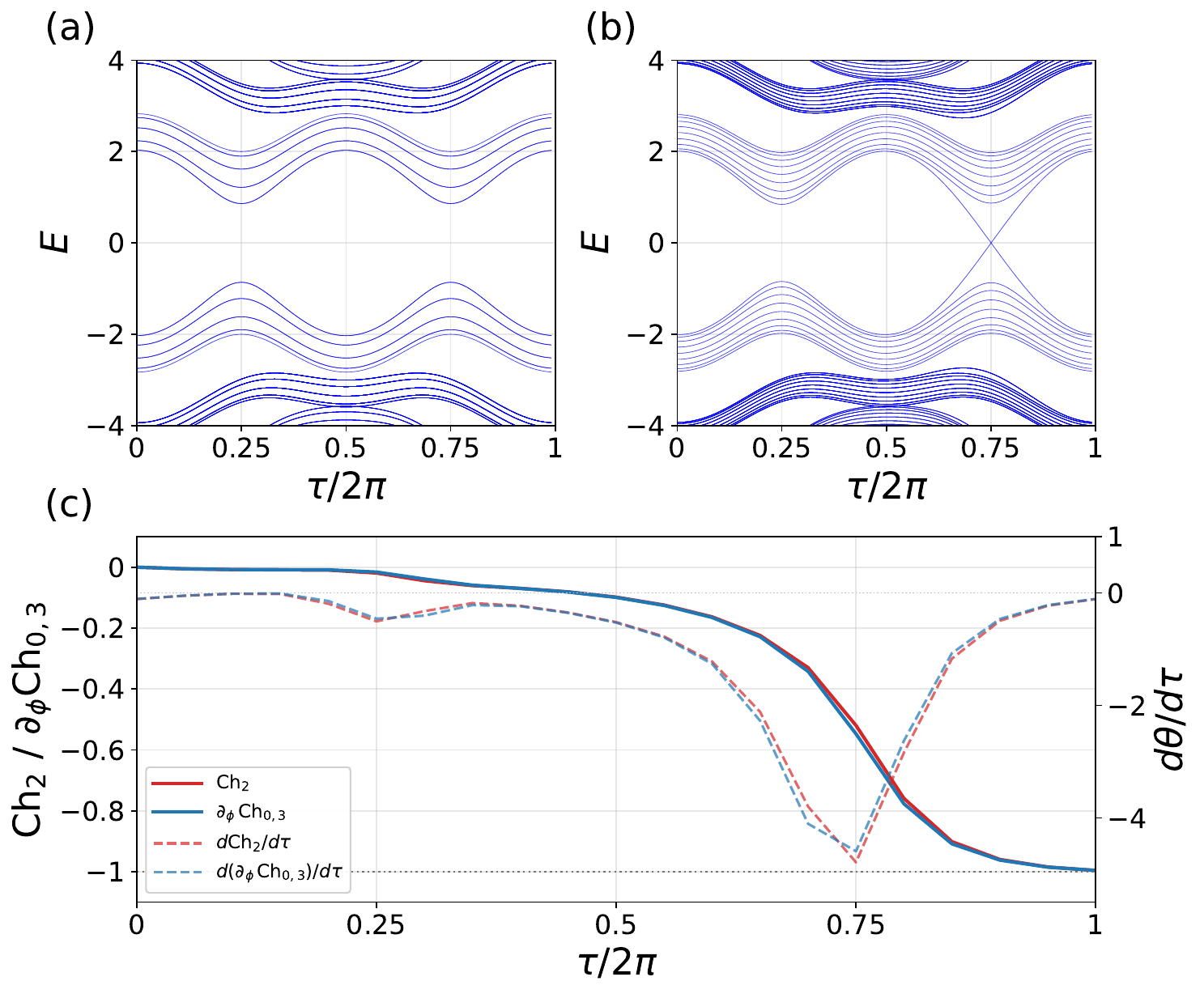}
\caption{Clean spectra and topological responses.
Panels~(a) and~(b) show spectra versus $\tau$ for $L=9$ with
periodic and open boundary conditions in~$z$, respectively.
Panel~(c) shows $\Chtwo(\tau)$ (red) and $\partial_\phi\Choz(\tau)$ (blue) at
$L=60$, $M=2048$, $N_\tau=20$; dashed curves show the $\tau$-resolved
integrands (right axis).}
\label{fig:clean}
\end{figure}

The second Chern number of the $(3{+}1)$D pump is formally expressed in real
space~\cite{Leung2013,Leung2020} as 
\begin{equation}
\Chtwo = 12\pi i \int_0^{2\pi} \! d\tau \;
\mathcal{T}\Bigl(
P \bigl[\partial_\tau P, [X_1, P]\bigr]
\bigl[X_2, P\bigr] \bigl[X_3, P\bigr] \Bigr),
\label{eq:ch2def}
\end{equation}
where $P = P(\tau)$ is the Fermi projection at $E_F=0$ and $X_j$ are
position operators.  For homogeneous disorder, translation covariance
identifies the trace per unit volume with the disorder-averaged local
matrix element at a reference orbital,
$\mathcal{T}(A)=\mathbb E \big [\langle 0|A|0\rangle\big ]$.  We
evaluate $\Chtwo$ on finite $L\times L\times L$ lattices with periodic
boundary conditions.  We approximate $P$ using a Chebyshev expansion in
$H$ of order~$M$ within the kernel polynomial method
(KPM)~\cite{Weisse2006}.  The commutators $[X_j,P]$ are regularized using
twisted projectors, while $\partial_\tau P$ is discretized over $N_\tau$
pump slices using finite differences~\cite{Prodan2017,SuppMat}.
Our implementation of the second Chern number reaches disordered
systems with up to $N = 4L^3 = 864{,}000$ orbitals, $N_\tau=20$, and
$M = 4096$,
making the mobility-gap regime accessible beyond exact
diagonalization~\cite{SuppMat}.  The trace per unit volume is strongly
self-averaging in the localized regime.  At $L = 60$, five disorder
realizations already yield standard errors
$\lesssim 0.001$--$0.006$ across all~$W$.

The bulk-boundary correspondence (BBC) proven in
Ref.~\cite{Leung2020} relates $\Chtwo$ to the surface spectral flow,
given by the net number of surface-state eigenvalues crossing~$E_F$
per pump cycle.  With perpendicular flux~$\phi_{12}$ applied through
the surface plane, surface states form Landau bands whose chiral flow
defines this spectral flow.  The BBC chain gives
$\mathrm{Sf}(\hat{h},G) = -\Choz$, where
$\mathrm{Sf}(\hat{h},G)$ denotes the spectral flow of the boundary
Hamiltonian $\hat{h}$ through the bulk gap~$G$, and the weak Chern
number measuring charge pumped along~$z$ is
\begin{equation}
\Choz = -2\pi i \int_0^{2\pi} \! d\tau \;
\mathcal{T}\bigl( P [\partial_\tau P, [X_3, P]] \bigr).
\label{eq:ch03}
\end{equation}
The generalized St\v{r}eda relation then yields
\begin{equation}
\partial_\phi \mathrm{Sf}(\hat{h},G)
= -\partial_\phi \Choz = -\Chtwo,
\label{eq:streda}
\end{equation}
where $\phi = \phi_{12}$ is the flux in units of the flux quantum.
Physically, $\Choz$ measures the net polarization charge pumped
along~$z$ per cycle, and $\partial_\phi \Choz$
measures how this pumped charge changes when a perpendicular magnetic
flux is applied.  In the surface picture of Ref.~\cite{Leung2020},
this flux quantizes the surface states into Landau bands, turning
each surface into a quantum Hall system.  The pump cycle drives
chiral flow of these Landau bands through~$E_F$, realizing a
surface Hall pumping process where the net spectral flow per cycle gives
the quantized Hall conductance carried by the surface.  The
derivative captures the rate at which this surface Hall
current changes per added flux quantum, directly linking the bulk
invariant $\Chtwo$ to the quantized Hall response of the pumped
surface.  Hence, we term $\partial_\phi \Choz$ the St\v{r}eda-Hall response.
$\partial_\phi \Choz$ is computed by inserting a uniform flux
$\phi = 1/L$ via Peierls phases and evaluated with the KPM
real-space method~\cite{SuppMat}.

As shown in Fig.~\ref{fig:clean}, the clean-limit PBC spectrum is
gapped, while OBC in~$z$ reveals surface states traversing the gap.
The real-space calculation gives
$\Chtwo=-0.996\pm0.001$, in agreement with the clean $k$-space result
of Ref.~\cite{Olsen2017}, and reproduces the generalized St\v{r}eda
relation at every pump slice.  The pronounced feature near
$\tau/2\pi\approx3/4$
coincides with the surface spectral flow point in
Fig.~\ref{fig:clean}(b).  The Fermi projector changes most rapidly
at the pump slice where a surface state eigenvalue crosses~$E_F$,
concentrating the integrand weight there.

\setcounter{topnumber}{1}
\begin{figure}[t]
\centering
\includegraphics[width=\columnwidth]{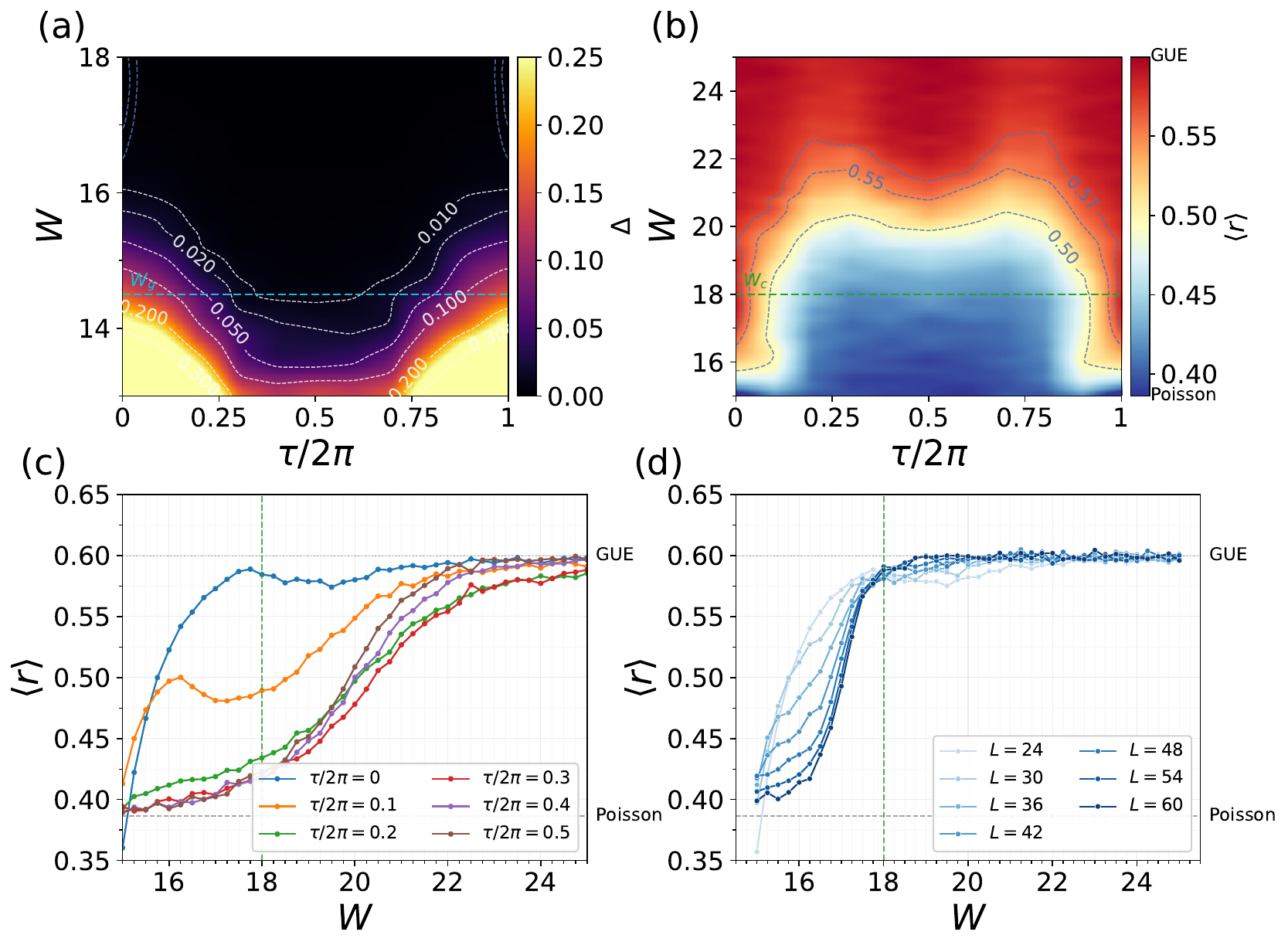}
\caption{Mobility-gap window at $E_F=0$.
(a)~Spectral gap for $L=60$ and 20 disorder realizations; the cyan line marks
global spectral gap closure $\Wg\approx14.5$, and blue dashed contours show
$\langle r\rangle=0.55$ and $0.57$ from panel~(b).
(b)~Level-spacing ratio $\langle r\rangle(\tau,W)$ for $L=24$
(500 realizations); dashed contours mark $0.50$, $0.55$, and $0.57$.
(c)~Fixed-$\tau$ cuts for $L=24$ ($k=30$, $|E|<0.25$, 1000 realizations).
(d)~Finite-size flow at $\tau=0$ for $L=24$--$60$
(1000 realizations at $L=24$, 500 otherwise), giving the critical disorder
$\Wc\approx18$.
See Ref.~\cite{SuppMat} for the $\tau$-resolved finite-size scaling.}
\label{fig:gap_lstats}
\end{figure}

The two disorder scales bounding the mobility-gap regime are distinct,
as shown in Fig.~\ref{fig:gap_lstats}.  The spectral gap and level
statistics use the eigenvalues of $H(\tau)$ closest to zero.
Defining the gap as $\Delta=\min(E>0)-\max(E<0)$, we identify spectral gap closure at $\Wg\approx14.5$, where the seed-averaged cycle
minimum $\Delta_{\min}(W)=\min_\tau\Delta(W,\tau)$ falls to
$\sim5\times10^{-3}$--$10^{-2}$.  Level statistics
use the adjacent-gap ratio
$r_n=\min(\delta_n,\delta_{n+1})/\max(\delta_n,\delta_{n+1})$,
averaged over near-zero eigenvalues and seeds.
Panel~(a) locates global spectral gap closure,
while panels~(b)--(d) identify and refine the critical disorder.  The
gap first vanishes near $\tau\approx\pi/2$, as shown in
Fig.~\ref{fig:gap_lstats}(a).  To locate the mobility-edge bottleneck,
we map the level-spacing ratio over the full pump cycle at $L=24$
[Fig.~\ref{fig:gap_lstats}(b)] and show corresponding fixed-$\tau$
slices in Fig.~\ref{fig:gap_lstats}(c).  These results identify
$\tau=0$ as the earliest-delocalizing slice.
The $\tau$-resolved level-statistics flows for $L=12$--$24$ in
Fig.~\ref{fig:r_fss}~\cite{SuppMat} show a Poisson-to-GUE crossover at
every sampled slice, consistent with a 3D unitary Anderson transition.
The targeted $\tau=0$ scaling in Fig.~\ref{fig:gap_lstats}(d), extended
to $L=60$, crosses near $W\approx18$, whereas the remaining sampled
slices, $\tau/2\pi=0.1$--$0.5$, cross only near
$W\approx19$--$20$.  We therefore estimate $W_c(0)\approx18$, which sets the
cycle-wide critical disorder $\Wc=\min_\tau W_c(\tau)\approx18$.

\begin{figure}[t]
\centering
\includegraphics[width=\columnwidth]{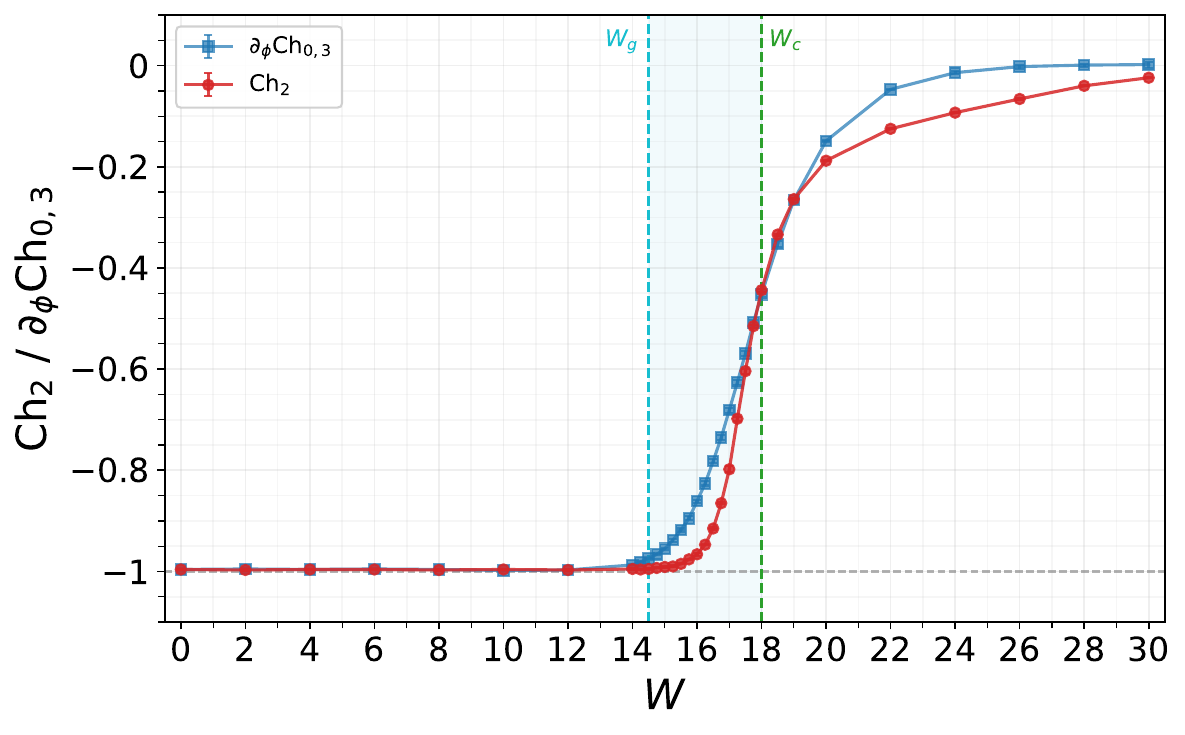}
\caption{$\Chtwo$ and $\partial_\phi\Choz$ versus $W$ at $L=60$,
$M=2048$ (5 disorder realizations).  The shaded mobility-gap regime extends from
$\Wg\approx14.5$ to $\Wc\approx18$.}
\label{fig:axion_surface_hall}
\end{figure}

The disorder dependence of $\Chtwo$ and the St\v{r}eda-Hall response
$\partial_\phi \Choz$ is shown in Fig.~\ref{fig:axion_surface_hall}
for $L = 60$ and $M = 2048$.  At weak disorder, both quantities form
a broad plateau at $-1$.  When the global spectral gap closes at
$\Wg\approx14.5$, the St\v{r}eda-Hall response begins to depart from the
plateau, while $\Chtwo$ remains closer to~$-1$ throughout much of the
mobility-gap regime.  Near $\Wc\approx18$, the plateau collapses as the
least-localized pump slice delocalizes, marking the disorder-driven
topological transition.  At stronger disorder, both quantities approach
zero, consistent with a topologically trivial insulating phase.

\begin{figure}[t]
\centering
\includegraphics[width=\columnwidth]{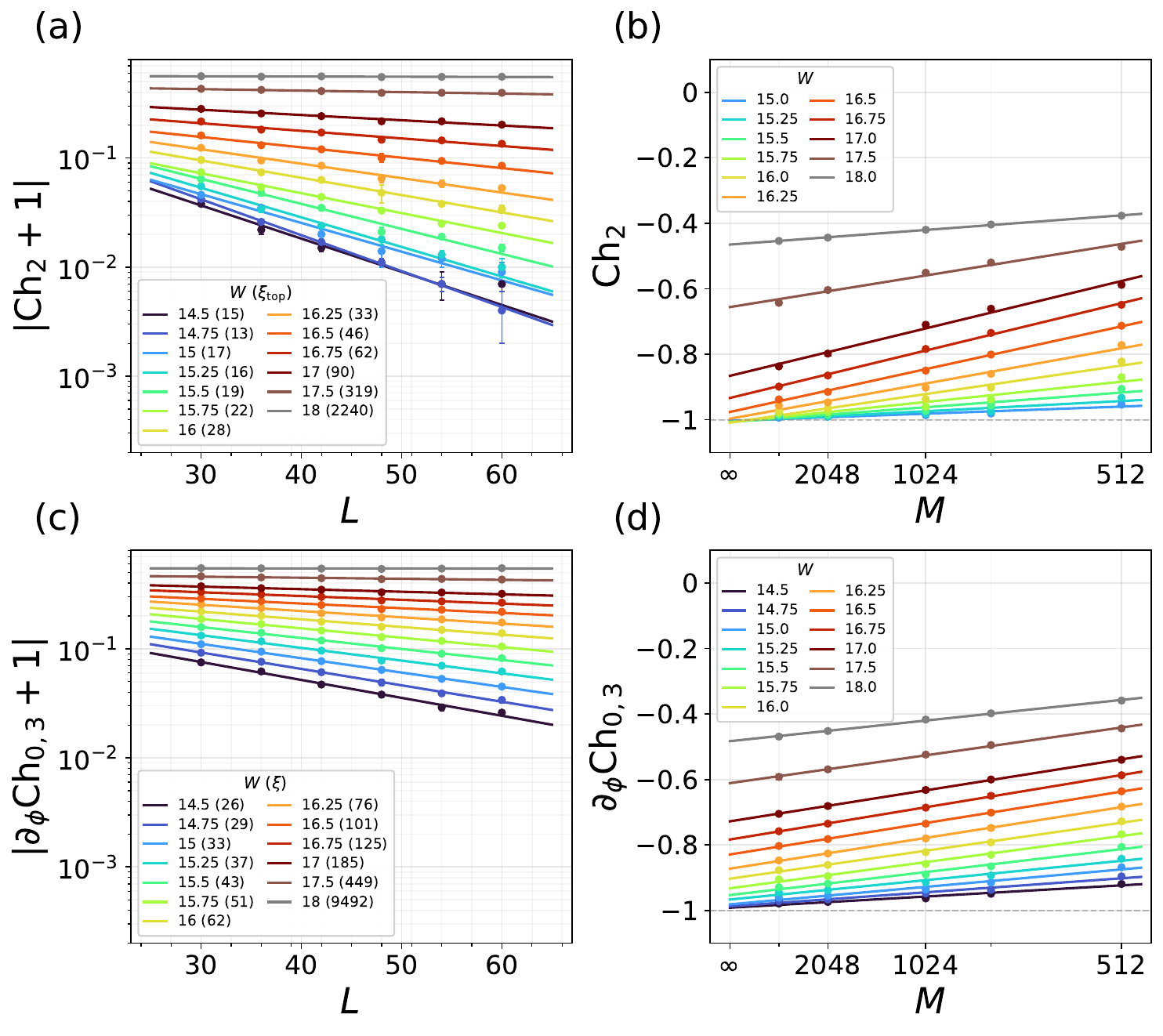}
\caption{Finite-size and KPM-order scaling for $W=14.5$--$18$.
Top: $\Chtwo$; bottom: $\partial_\phi\Choz$.
Panels~(a) and~(c) show deviations from~$-1$ versus~$L$, with exponential fits
and fit lengths $\xi_{\text{top}}$ and $\xi$, respectively
($M=2048$; 20 disorder realizations for $L=30$, 10 for $L=36,42$,
and 5 for $L=48,54,60$).
Panels~(b) and~(d) show KPM-order scaling at $L=60$ for
$M=512,768,1024,2048,4096$, using 5 disorder realizations, with linear fits in $1/M$.}
\label{fig:fss}
\end{figure}

We emphasize that the shading in Fig.~\ref{fig:axion_surface_hall}
denotes the cycle-wide mobility-gap regime, while
demonstrated $\Chtwo$ convergence extends through $W\lesssim16.5$.
The interval $16.5\lesssim W<\Wc$ remains unresolved because the
topological correction length exceeds accessible system sizes.  We characterize
this convergence by fitting
\begin{equation}
|\Chtwo(L; W) + 1| \sim A(W)e^{-L/\xi_{\text{top}}(W)},
\quad W < \Wc,
\label{eq:expdecay}
\end{equation}
where $\xi_{\text{top}}(W)$ is a topological correction length.
The fitted values of $\xi_{\text{top}}(W)$ are displayed in
parentheses in Fig.~\ref{fig:fss}(a).
Physically, $\xi_{\text{top}}$ measures the length scale over which
finite systems feel the approaching mobility edge.  When
$\xi_{\text{top}}\ll L$, the Fermi projector is effectively local on
the scale of the sample and the bulk invariant self-averages toward an
integer.  As $W$ approaches the mobility-edge bottleneck, $\xi_{\text{top}}$
grows; near $W\approx17$, it already exceeds $L_{\max}=60$, making the
extrapolation underconstrained although level statistics still support
localization.  Above $\Wc$, the $\tau=0$ slice
flows toward GUE statistics and $\Chtwo$ decays toward zero.
These results identify three distinct disorder scales: the global
spectral gap closes and the $\Chtwo$--$\partial_\phi\Choz$ separation begins at
$\Wg$, finite-size corrections become large near $W\approx17$, and
cycle-wide localization is lost at $\Wc$.

\setcounter{topnumber}{1}

The cumulative $\Chtwo$ and $\partial_\phi\Choz$ responses and their
$\tau$-resolved integrands in Fig.~\ref{fig:tau_resolved}
~\cite{SuppMat}
show that the separation between $\Chtwo$ and $\partial_\phi\Choz$ emerges
at global spectral gap closure, $\Wg\approx14.5$, and is concentrated
near the $\tau=0$ mobility-edge bottleneck and the surface spectral-flow
point at $\tau/2\pi=3/4$.  The persistence of this separation under
independent $L$- and $M$-scaling rules out finite flux and finite KPM
order as numerical artifacts.  The separation at $\Wg$ is
consistent with the global spectral-gap assumption underlying the
generalized St\v{r}eda relation, which no longer guarantees quantization
of $\partial_\phi\Choz$ once the spectral gap closes~\cite{ProdanSchulzBaldes2016}.
The $L$-scaling in Fig.~\ref{fig:fss}(a)
supports convergence of $\Chtwo$ toward~$-1$ according to
Eq.~\eqref{eq:expdecay}.  At $W = 16$, the fitted
$\xi_{\text{top}} \approx 28$ is well below $L_{\max}=60$, and the
fit is well converged.  As shown in Figs.~\ref{fig:fss}(a) and
\ref{fig:fss}(c), at $W=16$ and $L=60$, the deviations are
$|\Chtwo+1|=0.034$ and $|\partial_\phi\Choz+1|=0.139$, respectively.
Applying the same exponential fit to the St\v{r}eda-Hall response yields
a longer correction length, $\xi\approx62$, confirming that the response
does not converge to~$-1$ on the same scale.
Near $\Wc$, the fitted $\Chtwo$ correction length grows from
$\xi_{\text{top}}\approx90$ at $W=17$ to
$\xi_{\text{top}}\approx319$ at $W=17.5$ and
$\xi_{\text{top}}\approx2240$ at $W=18.0$, while the corresponding
St\v{r}eda-Hall correction lengths are even larger.  Thus, the
near-$\Wc$ data are finite-size limited rather than thermodynamically
saturated, consistent with a rapidly diverging mobility-bottleneck length.
The KPM-order scaling in panels~(b) and~(d) of Fig.~\ref{fig:fss}
further verifies that the early separation is not a KPM-resolution artifact.  At
$L=60$, linear extrapolation in $1/M$ using
$M\in\{512,768,1024,2048,4096\}$ gives $\Chtwo=-1.003$ and
$\partial_\phi\Choz=-0.981$ at $W=15.0$, a separation of $0.022$.
At $W=16.5$, the extrapolated values are $\Chtwo=-0.977$ and
$\partial_\phi\Choz=-0.829$, increasing the separation to $0.148$.
By $W=17.0$, both observables have moved appreciably away from
$-1$, with extrapolated values $\Chtwo=-0.866$ and
$\partial_\phi\Choz=-0.728$ and a separation of $0.138$.

\begin{figure}[t]
\centering
\includegraphics[width=\columnwidth]{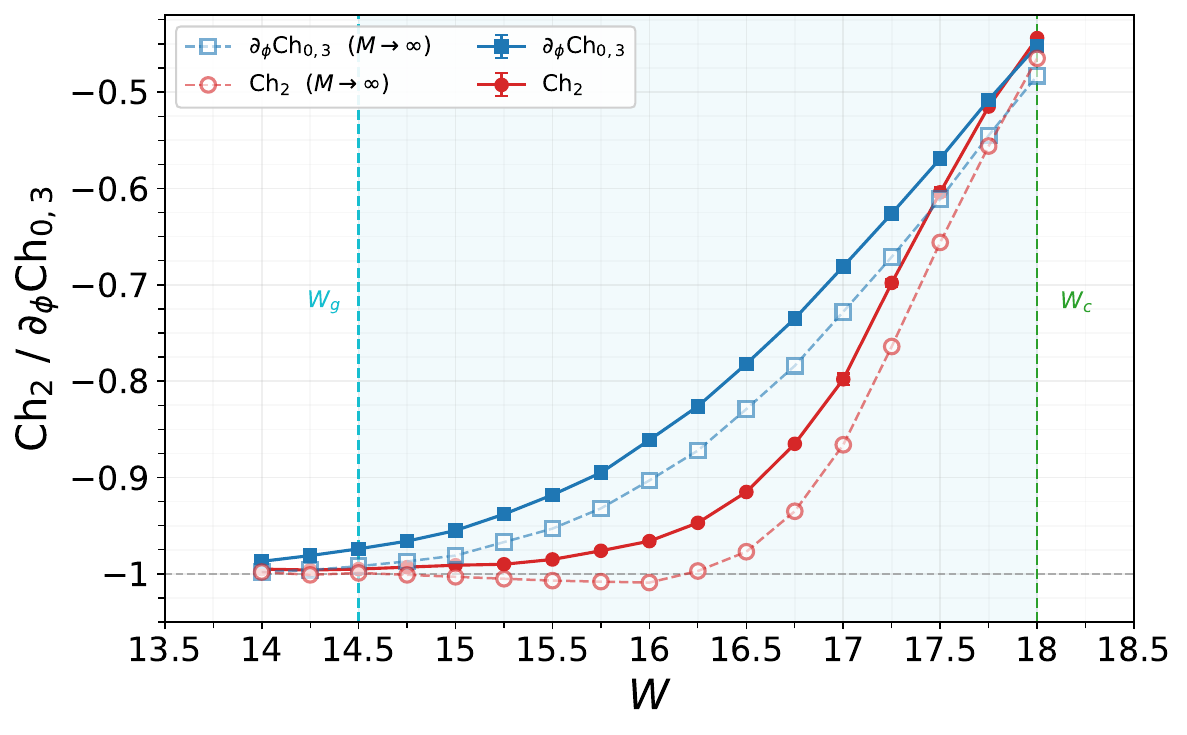}
\caption{KPM-order extrapolation at $L=60$ (5 disorder realizations).
Filled markers show $\Chtwo$ and $\partial_\phi\Choz$ at $M=2048$; open
markers show linear extrapolations to $1/M=0$.  Shading marks the
mobility-gap window.}
\label{fig:kpm_extrap}
\end{figure}

The full $M\to\infty$ extrapolated values of $\Chtwo$ and
$\partial_\phi\Choz$ are shown in Fig.~\ref{fig:kpm_extrap}.  Their
separation persists after extrapolation, ruling out finite KPM resolution
as its origin.  Beyond global spectral gap closure at $\Wg\approx14.5$,
the extrapolated $\Chtwo$ remains close to~$-1$ through $W\lesssim16.5$,
while $\partial_\phi\Choz$ departs more rapidly.  For
$W\gtrsim17$, both observables deviate appreciably from~$-1$; because
these results are obtained at fixed $L=60$, this regime remains finite-size
limited rather than demonstrating thermodynamic breakdown.  Together with
the $L$-scaling, the $M\to\infty$ extrapolation supports our central result:
$\Chtwo$ remains quantized well into the mobility-gap regime beyond global
spectral gap closure, while $\partial_\phi\Choz$ departs from quantization.

To conclude, we have provided substantial evidence for the quantization of the 3D axion pump beyond the spectral-gap regime. The standard proof of the quantization and invariance of the second Chern number in a mobility gap links $\Chtwo$ with the index of a properly chosen Fredholm operator~\cite{Prodan2013}. In the present context, this link could be established if the derivations $\nabla_\alpha P$ belong to a Sobolev space, \textit{i.e.}, if $\int d\tau \mathcal T(|\nabla_\alpha P|^4)<\infty$, where $\nabla_0=\partial_\tau$ and $\nabla_j=i[X_j,\mathord{\cdot}]$, $j=1,2,3$. Since $\tau$ lives on a circle, a Floquet transform can be used to place the model on a $\mathbb Z^4$ lattice and bring all coordinates on an equal footing. In a standard Anderson-localized regime on $\mathbb Z^4$, the Aizenman--Molchanov bound on the fractional moments of the resolvent can be used to bound these Sobolev norms~\cite{Aizenman1993,ETV2011,Aizenman1998,Prodan2013}.
However, a 3D axion pump differs in that there is no disorder in the zeroth direction. Under these conditions, estimating the Sobolev norm of $\partial_\tau P$, which encapsulates the required localization properties of the kernel $\mathbb E\big [\langle x|\partial_\tau P |y\rangle\big ]$, may require completely new techniques.

The quantization of the second Chern number of an axion pump is an important long-standing problem. A solution to this problem will finally confirm that the quantized magnetoelectric effect in 3D symmetry-protected topological insulators is robust beyond the spectral-gap regime~\cite{Leung2013,Prodan2013}. The breakdown reported for $(1{+}1)$D pumps shows that localization of the instantaneous states alone does not protect quantized pumping~\cite{Hayward2021}. By contrast, our numerical results provide strong evidence that a $(3{+}1)$D axion pump retains a quantized second Chern number in the mobility-gap regime. Numerically resolving this dimensional contrast between $(1{+}1)$D and $(3{+}1)$D pumps requires simulations at a substantially larger scale~\cite{SuppMat}.

\begin{acknowledgments} B.L. thanks David Vanderbilt for discussions. E.P. acknowledges support from the U.S. National Science Foundation through the grant CMMI-2131760.
\end{acknowledgments}

\clearpage
\onecolumngrid

\setcounter{figure}{0}
\renewcommand{\thefigure}{S\arabic{figure}}
\renewcommand{\theHfigure}{S\arabic{figure}}
\setcounter{table}{0}
\renewcommand{\thetable}{S\arabic{table}}
\renewcommand{\theHtable}{S\arabic{table}}
\setcounter{equation}{0}
\renewcommand{\theequation}{S\arabic{equation}}
\renewcommand{\theHequation}{S\arabic{equation}}

\begin{center}
{\large \textbf{Supplemental Material: Mobility-gap robustness and St\v{r}eda-Hall
separation in disordered axion pumping}}
\end{center}

\bigskip

\subsection{Real-space implementation of the second Chern number and St\v{r}eda-Hall response}

We evaluate the real-space second Chern number [Eq.~\eqref{eq:ch2def} of the main
text] on an $L \times L \times L$ lattice with periodic boundary
conditions.  The commutators $[X_j, P]$ are regularized using twisted
projectors~\cite{Prodan2017},
\begin{equation}
D_j P(\tau) = \sum_{m=-Q_s}^{Q_s} c_m^{(s)}\,
e^{-i m \Delta_j X_j}\,
P(\tau)\, e^{i m \Delta_j X_j},
\label{eq:twist}
\end{equation}
where $\Delta_j = 2\pi/L$ and the antisymmetric coefficients
$c_m^{(s)} = -c_{-m}^{(s)}$ are chosen so that
$D_j P \to -i[X_j, P]$ with error $O(\Delta_j^{2Q_s})$.  The
adiabatic derivative $\partial_\tau P$ is discretized independently,
\begin{equation}
D_0 P(\tau_l) = \sum_{m=-Q_\tau}^{Q_\tau} c_m^{(\tau)}\,
P(\tau_l + m \Delta_\tau),
\quad \Delta_\tau = \frac{2\pi}{N_\tau},
\label{eq:tauderiv}
\end{equation}
where $N_\tau=20$ is used.  The lattice
discretization can be written in terms of the trace per unit volume as
\begin{equation}
\Chtwo = \frac{-12\pi^2}{N_\tau} \sum_l
\mathcal{T}\!\left(P [D_0 P, D_1 P] [D_2 P, D_3 P]\right).
\label{eq:ch2realspace}
\end{equation}
In our calculation, $\mathcal{T}(A)$ is estimated using the stochastic trace as
$(L^3R)^{-1}\sum_{r=1}^{R}\langle r|A|r\rangle$
~\cite{Hutchinson1989}, where the number of independent random-phase
vectors is $R=15$.  For sufficiently localized operators, the
stochastic standard error is self-averaging and scales as
$\mathcal{O}[(RL^3)^{-1/2}]$~\cite{Weisse2006}.

Each unit cell contains four orbitals, so the Hilbert-space dimension is
$N=4L^3$.  Full dense diagonalization scales as
$\mathcal{O}(N^3)$ in time and
$\mathcal{O}(N^2)$ in memory.  At $L=60$,
$N=864{,}000$, making full diagonalization impractical; we therefore use
the kernel polynomial method (KPM)~\cite{Weisse2006}, which approximates the Fermi projector via a Chebyshev
expansion,
\begin{equation}
P \approx \sum_{n=0}^{M} \mu_n\, g_n\, T_n(\tilde{H}),
\label{eq:kpm}
\end{equation}
where $\tilde{H}$ is the Hamiltonian rescaled to $[-1,1]$, $T_n$ are
Chebyshev polynomials, $\mu_n$ are expansion coefficients for a step
function at the Fermi energy, and $g_n$ are Jackson damping factors.
The energy resolution is
$\delta E \sim \pi W_{\text{bw}}/M$, where $W_{\text{bw}}$ is the
bandwidth.  KPM reduces the per-sample complexity to $\mathcal{O}(MN)$,
linear in both the Chebyshev order $M$ and the system size $N$, with
$\mathcal{O}(N)$ memory per stochastic vector.  This linear scaling makes
it possible to average over disorder realizations and pump slices while
reaching the large system sizes needed to resolve the mobility-gap regime.

The weak Chern number $\Choz$, which measures the net charge pumped
along~$z$ per cycle, is computed analogously.  Its real-space formula
is
\begin{equation}
\Choz = -2\pi i \int_0^{2\pi} \! d\tau \;
\mathcal{T}\bigl( P [\partial_\tau P, [X_3, P]] \bigr),
\label{eq:ch03supp}
\end{equation}
which on the lattice becomes
\begin{equation}
\Choz = \frac{2\pi}{N_\tau} \sum_l
\mathcal{T}\!\left(P [D_0 P, D_3 P]\right),
\label{eq:ch03lattice}
\end{equation}
using the spatial and temporal discretizations in
Eqs.~\eqref{eq:twist} and \eqref{eq:tauderiv} for $D_3$ and $D_0$,
respectively.  The
St\v{r}eda-Hall response is then obtained by computing $\Choz$ at two
flux values and forming the finite difference
\begin{equation}
\partial_\phi \Choz \approx
\frac{\Choz(\phi) - \Choz(0)}{\phi}, \quad
\phi = \frac{1}{L}.
\label{eq:streda_fd}
\end{equation}
The magnetic flux $\phi = 1/L$ (one flux quantum per $L$ plaquettes)
is introduced through the $x$-$y$ plane via Peierls phases on the
hopping terms,
$t_{ij} \to t_{ij} \exp(2\pi i \phi \langle \bm{r}_i,
\Phi \bm{r}_j \rangle)$, where $\Phi$ is the antisymmetric flux
matrix with $\Phi_{12} = -\Phi_{21} = 1/2$.
This orientation gives the sign convention used in
Eq.~\eqref{eq:streda} of the main text.

\subsection{Noncommutative first Chern number and quantized pump}

Here, we compute the noncommutative first Chern number of the disordered Rice--Mele model studied in
Ref.~\cite{Hayward2021}.  The model is a one-dimensional chain with
periodic boundary conditions, alternating hoppings
$J[1 \pm \delta(\tau)]$, staggered on-site potential
$\Delta(\tau)$, and uniform diagonal disorder
$V_j \in [-W/2,W/2]$.  We use the pump cycle
\begin{equation}
\delta(\tau) = R_\delta \cos\tau, \qquad
\Delta(\tau) = R_\Delta \sin\tau,
\end{equation}
with $R_\delta = 0.5$, $R_\Delta = 2.3$, and $J=1$.  This gives a
clean Thouless pump~\cite{Thouless1983} with $|\Chone| = 1$.

\begin{figure}[!b]
\centering
\includegraphics[width=0.70\textwidth]{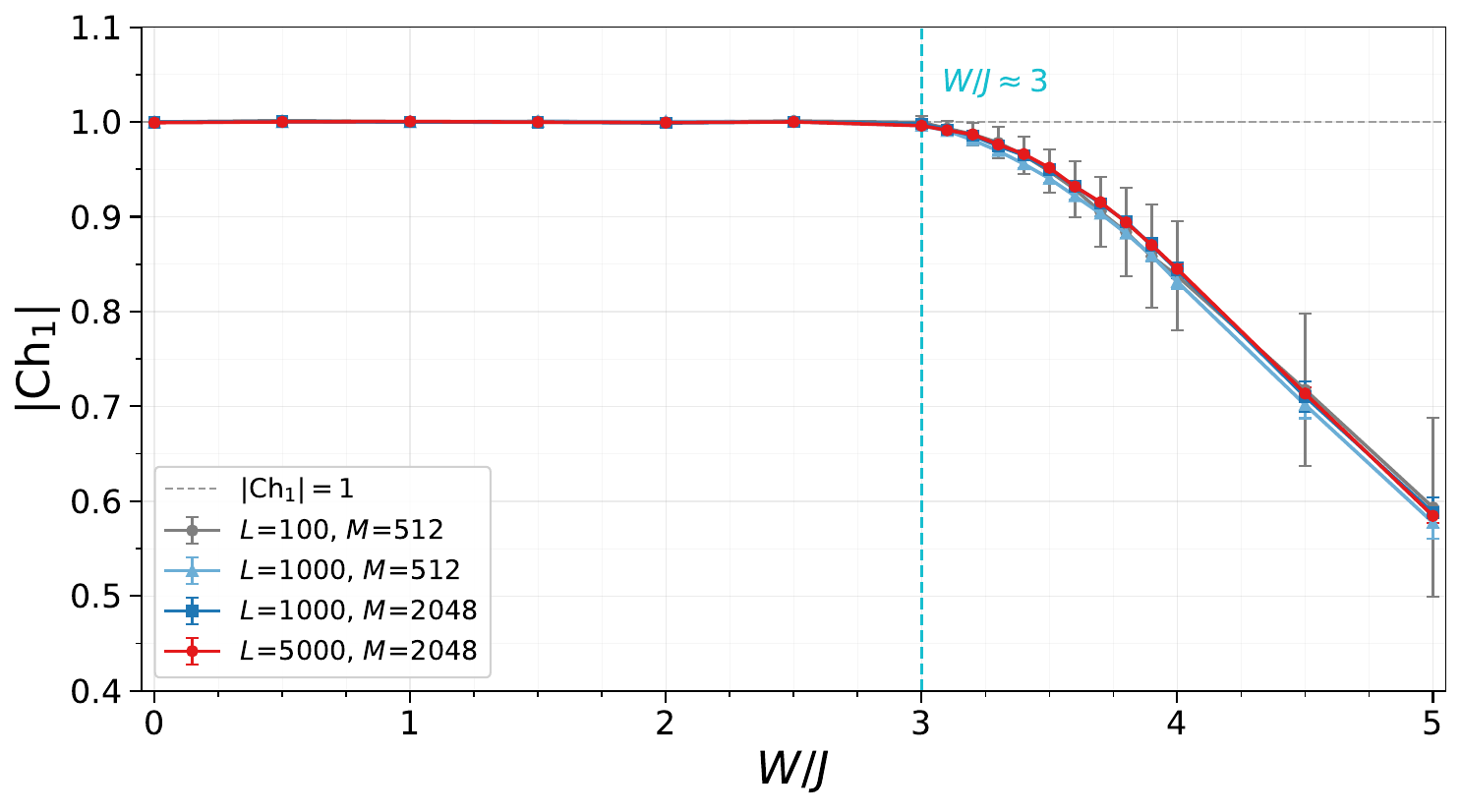}
\caption{Noncommutative first Chern number $\Chone$ versus $W/J$ for the disordered
Rice--Mele pump.  $|\Chone|$ remains quantized at weak
disorder and decreases near spectral gap closure at $W/J\approx3$.
All curves use $N_\tau=100$, with 10 disorder
realizations for $L=100$ and 5 for $L=1000,5000$.}
\label{fig:rice_mele_kpm}
\end{figure}

The noncommutative first Chern number governing the pumped charge is
\begin{equation}
\Chone = -2\pi i \int_0^{2\pi} \! d\tau \;
\mathcal{T}\bigl(P[\partial_\tau P,[X,P]]\bigr),
\label{eq:rice_mele_c1_continuum}
\end{equation}
where $\mathcal{T}$ is the trace per unit length.  Here $D_1P$ is the
twisted finite-difference derivative along the chain and $D_0P$ is
the finite-difference derivative along the pump coordinate~$\tau$.
With the same discretization used above, the lattice expression is
\begin{equation}
\Chone = \frac{2\pi}{N_\tau} \sum_l
\mathcal{T}\!\left(P [D_0 P, D_1 P]\right),
\label{eq:rice_mele_c1}
\end{equation}
where $\Chone$ is the noncommutative first Chern number and its disorder
dependence reproduces the breakdown of quantized pumping reported in
Fig.~12 of Ref.~\cite{Hayward2021}.
The real-space Rice--Mele calculations in Fig.~\ref{fig:rice_mele_kpm}
show that $|\Chone|$ remains quantized at weak disorder and decreases
after the spectral-gap-closure scale near $W/J \approx 3$,
reproducing the reported results.
Increasing the system size and KPM order leaves the downturn intact,
indicating loss of quantized pumping rather than a finite-size or
energy-resolution artifact.  Thus, in one dimension, localization of
the instantaneous states does not by itself protect the pumped Chern
number after spectral gap closure.
This contrasts with the 3D axion pump in
the main text, where the bulk second Chern number remains close to
its quantized value beyond global spectral gap closure in the mobility-gap
regime.

\FloatBarrier

\subsection{\texorpdfstring{$\tau$-resolved level statistics}{tau-resolved level statistics}}

\begin{figure}[!htbp]
\centering
\includegraphics[width=\textwidth]{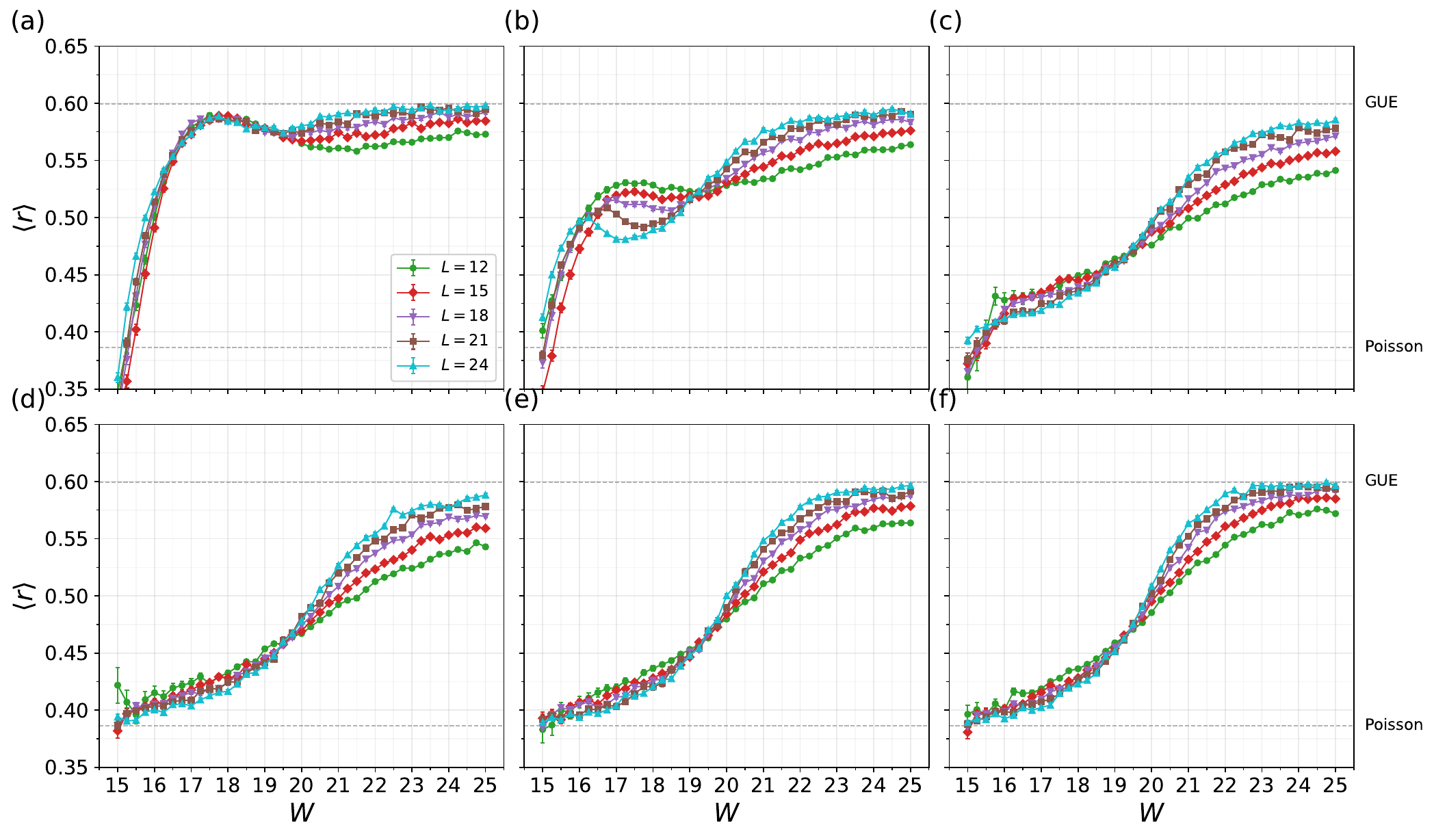}
\caption{Level-spacing ratio $\langle r\rangle$ versus $W$ for
$\tau/2\pi=0,0.1,0.2,0.3,0.4,0.5$ in panels~(a)--(f), respectively
($L=12$--$24$, $|E|<0.25$, 1000 disorder realizations).}
\label{fig:r_fss}
\end{figure}

The $\tau$-resolved level-statistics flows in Fig.~\ref{fig:r_fss} are
computed using shift-invert Lanczos near $E_F=0$ for six pump slices in
the first half-cycle, $\tau/2\pi=0$--$0.5$, using $L=12$--$24$.
The full-cycle map in Fig.~\ref{fig:gap_lstats}(b) is approximately
symmetric under $\tau/2\pi\to1-\tau/2\pi$, so these slices capture the
representative cycle dependence.  These data establish
the slice dependence of $W_c(\tau)$ and identify $\tau=0$ as the
mobility-edge bottleneck.  The targeted calculation in
Fig.~\ref{fig:gap_lstats}(d) of the main text subsequently extends this slice to
$L=60$ and refines its critical disorder to
$W_c(0)\approx18$.  Other slices cross only near
$W\approx19$--$20$, so the $\tau=0$ transition determines the loss
of cycle-wide localization.

\FloatBarrier

\subsection{\texorpdfstring{$\tau$-resolved $\Chtwo$--$\partial_\phi\Choz$ separation}{tau-resolved Ch2--partial-phi Ch0,3 separation}}

\begin{figure}[!htbp]
\centering
\includegraphics[width=\textwidth]{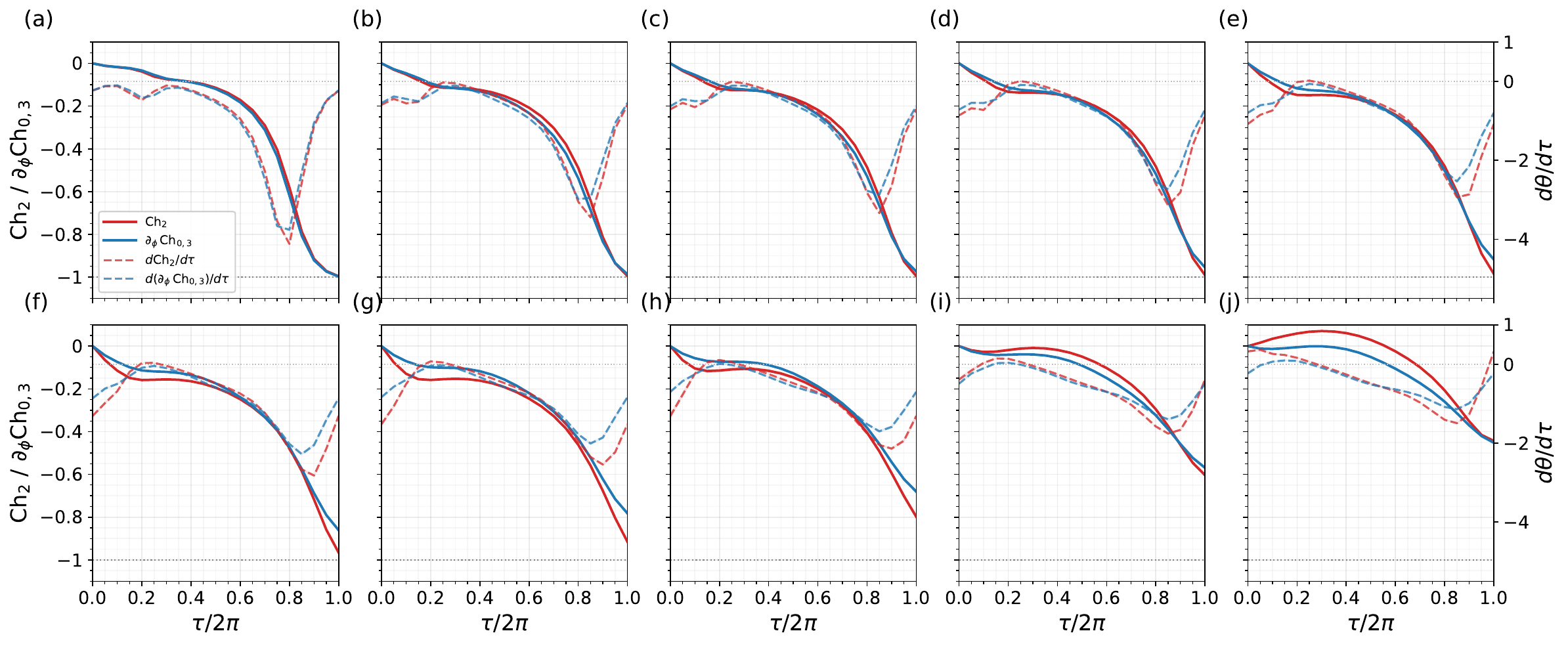}
\caption{The $\tau$-resolved $\Chtwo$ and $\partial_\phi\Choz$ responses are shown for
$W=10,14,14.5,15,15.5,16,16.5,17,17.5,18$ in panels~(a)--(j), respectively
($L=60$, $M=2048$, $N_\tau=20$).  Solid lines show the cumulative responses,
and dashed lines show their $\tau$-resolved integrands (right axis).
Shaded bands indicate standard errors over 5 disorder realizations.}
\label{fig:tau_resolved}
\end{figure}

Figure~\ref{fig:tau_resolved} compares the cumulative integrals and
$\tau$-resolved integrands of $\Chtwo$ and
$\partial_\phi\Choz$ for $W=10$--$18$ ($L=60$, $M=2048$).  Below the
global spectral gap closure at $\Wg\approx14.5$, the two integrands
coincide.  Above $\Wg$, their separation concentrates near $\tau=0$ and
$\tau/2\pi \approx 3/4$, consistent with the mobility-edge bottleneck and
spectral flow sectors identified in the main text.
Although $\Chtwo$ remains closer to its quantized value of~$-1$ than
$\partial_\phi\Choz$ for $W \lesssim 17$, it also deviates strongly near
$W \approx 18$.

\FloatBarrier


\begin{thebibliography}{0}%
\makeatletter
\providecommand \@ifxundefined [1]{%
 \@ifx{#1\undefined}
}%
\providecommand \@ifnum [1]{%
 \ifnum #1\expandafter \@firstoftwo
 \else \expandafter \@secondoftwo
 \fi
}%
\providecommand \@ifx [1]{%
 \ifx #1\expandafter \@firstoftwo
 \else \expandafter \@secondoftwo
 \fi
}%
\providecommand \natexlab [1]{#1}%
\providecommand \enquote  [1]{``#1''}%
\providecommand \bibnamefont  [1]{#1}%
\providecommand \bibfnamefont [1]{#1}%
\providecommand \citenamefont [1]{#1}%
\providecommand \href@noop [0]{\@secondoftwo}%
\providecommand \href [0]{\begingroup \@sanitize@url \@href}%
\providecommand \@href[1]{\@@startlink{#1}\@@href}%
\providecommand \@@href[1]{\endgroup#1\@@endlink}%
\providecommand \@sanitize@url [0]{\catcode `\\12\catcode `\$12\catcode
  `\&12\catcode `\#12\catcode `\^12\catcode `\_12\catcode `\%12\relax}%
\providecommand \@@startlink[1]{}%
\providecommand \@@endlink[0]{}%
\providecommand \url  [0]{\begingroup\@sanitize@url \@url }%
\providecommand \@url [1]{\endgroup\@href {#1}{\urlprefix }}%
\providecommand \urlprefix  [0]{URL }%
\providecommand \Eprint [0]{\href }%
\providecommand \doibase [0]{https://doi.org/}%
\providecommand \selectlanguage [0]{\@gobble}%
\providecommand \bibinfo  [0]{\@secondoftwo}%
\providecommand \bibfield  [0]{\@secondoftwo}%
\providecommand \translation [1]{[#1]}%
\providecommand \BibitemOpen [0]{}%
\providecommand \bibitemStop [0]{}%
\providecommand \bibitemNoStop [0]{.\EOS\space}%
\providecommand \EOS [0]{\spacefactor3000\relax}%
\providecommand \BibitemShut  [1]{\csname bibitem#1\endcsname}%
\let\auto@bib@innerbib\@empty
\end{thebibliography}%


\begin{thebibliography}{99}

\bibitem{Qi2008}
X.-L.~Qi, T.~L.~Hughes, and S.-C.~Zhang,
Phys.\ Rev.\ B \textbf{78}, 195424 (2008).

\bibitem{Essin2009}
A.~M.~Essin, J.~E.~Moore, and D.~Vanderbilt,
Phys.\ Rev.\ Lett.\ \textbf{102}, 146805 (2009).

\bibitem{Mong2010}
R.~S.~K.~Mong, A.~M.~Essin, and J.~E.~Moore,
Phys.\ Rev.\ B \textbf{81}, 245209 (2010).

\bibitem{Leung2020}
B.~Leung and E.~Prodan,
J.\ Phys.\ A: Math.\ Theor.\ \textbf{53}, 205203 (2020).

\bibitem{Leung2013}
B.~Leung and E.~Prodan,
J.\ Phys.\ A: Math.\ Theor.\ \textbf{46}, 085205 (2013).

\bibitem{Prodan2013}
E.~Prodan, B.~Leung, and J.~Bellissard,
J.\ Phys.\ A: Math.\ Theor.\ \textbf{46}, 485202 (2013).

\bibitem{Wu2016}
L.~Wu \emph{et al.},
Science \textbf{354}, 1124 (2016).

\bibitem{Mogi2017}
M.~Mogi \emph{et al.},
Sci.\ Adv.\ \textbf{3}, eaao1669 (2017).

\bibitem{Xiao2018}
D.~Xiao \emph{et al.},
Phys.\ Rev.\ Lett.\ \textbf{120}, 056801 (2018).

\bibitem{Liu2020}
C.~Liu \emph{et al.},
Nat.\ Mater.\ \textbf{19}, 522 (2020).

\bibitem{LiuGap2022}
M.~Liu \emph{et al.},
Proc.\ Natl.\ Acad.\ Sci.\ U.S.A.\ \textbf{119}, e2207681119 (2022).

\bibitem{Zhuo2023}
D.~Zhuo \emph{et al.},
Nat.\ Commun.\ \textbf{14}, 7596 (2023).

\bibitem{Qiu2025}
J.-X.~Qiu \emph{et al.},
Nature \textbf{641}, 62 (2025).

\bibitem{Hu2026}
J.~Hu \emph{et al.},
Nat.\ Commun.\ \textbf{17}, 1305 (2026).

\bibitem{NomuraNagaosa2011}
K.~Nomura and N.~Nagaosa,
Phys.\ Rev.\ Lett.\ \textbf{106}, 166802 (2011).

\bibitem{LeungProdan2012}
B.~Leung and E.~Prodan,
Phys.\ Rev.\ B \textbf{85}, 205136 (2012).

\bibitem{Li2021}
H.~Li, H.~Jiang, C.-Z.~Chen, and X.~C.~Xie,
Phys.\ Rev.\ Lett.\ \textbf{126}, 156601 (2021).

\bibitem{Song2021}
Z.-D.~Song, B.~Lian, R.~Queiroz, R.~Ilan, B.~A.~Bernevig, and A.~Stern,
Phys.\ Rev.\ Lett.\ \textbf{127}, 016602 (2021).

\bibitem{Chen2025STI}
X.~Chen, F.-J.~Wang, Z.~Bi, and Z.-D.~Song,
Phys.\ Rev.\ Lett.\ \textbf{134}, 226601 (2025).

\bibitem{Grindall2025}
C.~Grindall, A.~C.~Tyner, A.-K.~Wu, T.~L.~Hughes, and J.~H.~Pixley,
Phys.\ Rev.\ Lett.\ \textbf{135}, 226601 (2025).

\bibitem{Bellissard1994}
J.~Bellissard, A.~van Elst, and H.~Schulz-Baldes,
J.\ Math.\ Phys.\ \textbf{35}, 5373 (1994).

\bibitem{Thouless1983}
D.~J.~Thouless,
Phys.\ Rev.\ B \textbf{27}, 6083 (1983).

\bibitem{Wauters2019}
M.~M.~Wauters, A.~Russomanno, R.~Citro, G.~E.~Santoro, and L.~Privitera,
Phys.\ Rev.\ Lett.\ \textbf{123}, 266601 (2019).

\bibitem{Cerjan2020}
A.~Cerjan, M.~Wang, S.~Huang, K.~P.~Chen, and M.~C.~Rechtsman,
Light Sci.\ Appl.\ \textbf{9}, 178 (2020).

\bibitem{Hayward2021}
A.~L.~C.~Hayward, E.~Bertok, U.~Schneider, and
F.~Heidrich-Meisner,
Phys.\ Rev.\ A \textbf{103}, 043310 (2021).

\bibitem{Huang2025}
Y.~Liu \emph{et al.},
Nat.\ Commun.\ \textbf{16}, 108 (2025).

\bibitem{SuppMat}
See Supplemental Material for implementation details, convergence
analyses, and level statistics.

\bibitem{Abrahams1979}
E.~Abrahams, P.~W.~Anderson, D.~C.~Licciardello, and
T.~V.~Ramakrishnan,
Phys.\ Rev.\ Lett.\ \textbf{42}, 673 (1979).

\bibitem{ProdanSchulzBaldes2016}
E.~Prodan and H.~Schulz-Baldes,
\emph{Bulk and Boundary Invariants for Complex Topological Insulators:
From K-Theory to Physics} (Springer, 2016).

\bibitem{Streda1982}
P.~St\v{r}eda,
J.\ Phys.\ C \textbf{15}, L717 (1982).

\bibitem{Olsen2017}
T.~Olsen, M.~Taherinejad, D.~Vanderbilt, and I.~Souza,
Phys.\ Rev.\ B \textbf{95}, 075137 (2017).

\bibitem{Weisse2006}
A.~Wei{\ss}e, G.~Wellein, A.~Alvermann, and H.~Fehske,
Rev.\ Mod.\ Phys.\ \textbf{78}, 275 (2006).

\bibitem{Prodan2017}
E.~Prodan,
\emph{A Computational Non-commutative Geometry Program for Disordered
Topological Insulators} (Springer, 2017).

\bibitem{Aizenman1993}
M.~Aizenman and S.~Molchanov,
Commun.\ Math.\ Phys.\ \textbf{157}, 245 (1993).

\bibitem{ETV2011}
A.~Elgart, M.~Tautenhahn, and I.~Veseli{\'c},
Ann.\ Henri Poincar{\'e} \textbf{12}, 1571 (2011).

\bibitem{Aizenman1998}
M.~Aizenman and G.~M.~Graf,
J.\ Phys.\ A: Math.\ Gen.\ \textbf{31}, 6783 (1998).

\bibitem{Hutchinson1989}
M.~F.~Hutchinson,
Commun.\ Stat.\ Simul.\ Comput.\ \textbf{18}, 1059 (1989).

\end{thebibliography}
\end{document}